\documentclass[aps, prl, superscriptaddress, reprint]{revtex4-1}

\usepackage{siunitx}
\usepackage{chemformula}
\usepackage{amsmath}
\usepackage{amssymb}
\usepackage{graphicx}
\usepackage{hyperref}
\usepackage{verbatim}
\usepackage{lipsum}

\newcommand{\mr}[1]{\mathrm{#1}}

\newcommand{\cf}[0]{cf.~}
\newcommand{\ie}[0]{i.e.,~}

\newcommand{\Fref}[1]{Figure~\ref{fig:#1}}
\newcommand{\fref}[1]{Fig.~\ref{fig:#1}}

\newcommand{\eref}[1]{Eq.~(\ref{eq:#1})}
\newcommand{\Cref}[1]{Chapter~\ref{chap:#1}}
\newcommand{\cref}[1]{Ch.~\ref{chap:#1}}

\newcommand{\rs}[1]{{\color{black}#1}}

\renewcommand{\eth}[0]{Department of Materials, ETH Z{\"u}rich, 8093 Z{\"u}rich, Switzerland}

\newcommand{\tud}[0]{Institut f{\"u}r Festk{\"o}rper- und Materialphysik, Technische Universit{\"a}t Dresden and W{\"u}rzburg-Dresden Cluster of Excellence ct.qmat, 01062 Dresden, Germany}
\newcommand{\ifw}[0]{Leibniz Institute for Solid State and Materials Research Dresden (IFW Dresden), Institute for Metallic Materials, 01069 Dresden, Germany}

\newcommand{\ukn}[0]{Department of Physics, University of Konstanz, 78457 Konstanz, Germany}
\newcommand{\ntnu}[0]{Center for Quantum Spintronics, Department of Physics, Norwegian University of Science and Technology, 7491 Trondheim, Norway}
\newcommand{\ifimac}[0]{Condensed Matter Physics Center (IFIMAC) and Departamento de F\'{i}sica Te\'{o}rica de la Materia Condensada, Universidad Aut\'{o}noma de Madrid, E-28049 Madrid, Spain}

\begin{document}

\title{Control of nonlocal magnon spin transport via magnon drift currents}

\author{Richard Schlitz}
\email{richard.schlitz@mat.ethz.ch}
\affiliation{\eth}
\affiliation{\tud}
\author{Sa\"ul V\'elez}
\email{saul.velez@mat.ethz.ch, Present address: Condensed Matter Physics Center (IFIMAC) and Departamento de F\'{i}sica de la Materia Condensada, Universidad Aut\'{o}noma de Madrid, E-28049 Madrid, Spain}
\affiliation{\eth}
\author{Akashdeep Kamra}
\affiliation{\ntnu}
\affiliation{\ifimac}
\author{Charles-Henri Lambert}
\affiliation{\eth}
\author{Michaela Lammel}
\affiliation{\ifw}
\affiliation{\ukn}
\author{Sebastian T. B. Goennenwein}
\affiliation{\tud}
\affiliation{\ukn}
\author{Pietro Gambardella}
\affiliation{\eth}

\date{\today}

\begin{abstract}
    Spin transport via magnon diffusion in magnetic insulators is important for a broad range of spin-based phenomena and devices. However, the absence of the magnon equivalent of an electric force is a bottleneck. In this work, we demonstrate the controlled generation of magnon drift currents in yttrium iron garnet/platinum heterostructures. By performing electrical injection and detection of incoherent magnons, we find magnon drift currents that stem from the interfacial Dzyaloshinskii-Moriya interaction. We can further control the magnon drift by the orientation of the magnetic field. The drift current changes the magnon propagation length by up to $\pm\SI{6}{\percent}$ relative to diffusion. We generalize the magnonic spin transport theory to include a finite drift velocity resulting from any inversion asymmetric interaction, and obtain results consistent with our experiments.
\end{abstract}

\maketitle


\emph{Introduction.} Magnons are collective excitations of ordered magnets that transport spin information without any associated charge motion, unlike electrons.
This places magnons and magnon spin currents at the core of an emerging spin-based paradigm for information transport and processing~\cite{Kruglyak2010, Kajiwara2010, Chumak2015, Kikkawa2016, Chumak2017a, Lebrun2018, Harii2019, Cornelissen2015}. 
In a prototypical device~\cite{Cornelissen2015, Zhang2012, Zhang2012a, Takei2014,Goennenwein2015,Cornelissen2016a,Li2016}, magnons are injected and detected electrically via the spin Hall effect in two electrically independent but closely spaced heavy metal wires on a magnetic insulator~\cite{Dyakonov1971, Hirsch1999,Valenzuela2006,Saitoh2006,Sinova2015}. 
A broad range of devices allowing information transport and logic operations - in analogy to electronic devices - have already been accomplished within this paradigm of magnonics~\cite{Chumak2014,Ganzhorn2016a,Shan2017,Lebrun2018,Cornelissen2018, Oyanagi2019,  Ross2019, Wimmer2019, Avci2020, Wimmer2020, GomezPerez2020}. 
In electronics, an electric force on the charge carriers is straightforwardly generated by electric fields.
The ensuing electronic drift currents play the central role in various electronic devices~\cite{Zutic2004, Manchon2019}. 
In contrast, magnetic fields \rs{do not} exert forces on magnons since they do not provide the required inversion symmetry breaking and thus cannot generate magnon drift currents.
The generation of magnon drift currents promises to expand the functionality of magnonic devices.

Magnets breaking inversion symmetry allow for an antisymmetric exchange or Dzyaloshinskii-Moriya interaction~\cite{Dzyaloshinsky1958,Moriya1960} (DMI)\rs{, which is responsible for the emergence of }topological spin textures~\cite{Gobel2020}, the efficient driving of domain walls by spin-orbit torques~\cite{Velez2019a, Avci2019} and the existence of non-reciprocal spin wave modes~\cite{Wang2020}. 
The DMI can either stem from the bulk crystal structure~\cite{Dzyaloshinsky1958, Moriya1960} or from \rs{interface-induced symmetry breaking in heterostructures including magnetic and nonmagnetic materials}~\cite{Miyawaki2017, Fert2017, Bode2007, Ryu2013}.
As a result, engineering the DMI in \rs{magnetic} heterostructures has become a key focus of research~\cite{Bode2007, Ryu2013, Emori2013, Chen2013a, Nembach2015, Ding2019, Avci2019, Velez2019a, Wang2020}. 


In this Letter, we show that the DMI leads to a finite magnon drift velocity superposed on diffusive spin transport mediated by incoherent magnons. 
We study nonlocal magnon transport in devices with three parallel, electrically independent Pt wires on yttrium iron garnet (YIG).
When driving an electric current through the central injector wire, the interfacial spin accumulation arising from the spin Hall effect excites magnons below the injector (\fref{f1}a).
In the absence of a magnon drift velocity, an equal amount of magnons diffuse towards the two Pt wires on either side of the injector, leading to identical voltages generated via the inverse spin Hall in either wire.
A finite magnon drift velocity \rs{due to the} interfacial DMI, however, leads to an asymmetric propagation of the injected magnons, resulting in a larger electrical signal on the left wire compared to the right one (\fref{f1}a). 
We show that such a directional transport contribution is present in our YIG/Pt heterostructures and elucidate its interfacial origin from YIG thickness dependent measurements.
We generalize the diffusive magnon transport theory to include inversion asymmetry and find good agreement with our experimental observations.
Our work shows that nonlocal magnon transport experiments can be used to probe the presence of DMI in magnetic insulators.


\emph{Magnon dispersion with DMI.} In the presence of DMI and in the limit of small magnon wave vectors $\mathbf{k}$, the magnon dispersion relation $\omega_\mr{s} \propto \mathbf{k}^2$ is superposed to an asymmetric contribution $\omega_{\mr{a}} \propto \mathbf{k}$, the slope of which corresponds to the magnon drift velocity $\mathbf{v}_\mr{DMI}$~\cite{Moon2013, Nembach2015, Wang2020} (\cf \fref{f1}b).
For interfacial DMI with symmetry breaking along $\hat{\mathbf{z}}$~\cite{Moon2013, Wang2020}
\begin{equation}\label{eq:vdrift}
    \omega_\mr{a} = \mathbf{v}_\mr{DMI} \cdot \mathbf{k} = [(\hat{\mathbf{z}} \times \hat{\mathbf{m}})\cdot \mathbf{k}] \frac{2 \gamma}{M_\mr{s}} D,
\end{equation}
where $\omega$ is the angular frequency of the magnon, $\gamma > 0$ the gyromagnetic ratio, $M_\mr{s}$ the saturation magnetization, and $D$ the constant quantifying the sign and strength of the DMI.
Consequently, the orientation of the drift velocity $\mathbf{v}_\mr{DMI}$ can be controlled by orienting the magnetization direction $\hat{\mathbf{m}} = \mathbf{M}/M_\mr{s}$~\cite{Udvardi2009, Moon2013, Nembach2015} (\cf \fref{f1}b).
For magnon transport along $\hat{\mathbf{y}}$ and when the angle of the magnetic field $\mathbf{H}$ (and thus of $\hat{\mathbf{m}}$) with respect to the current direction $\hat{\mathbf{x}}$ is given by $\alpha$, we find $\left.\mathbf{v}_\mr{DMI}\right|_{\hat{\mathbf{y}}} = v_{\mr{DMI}} \cos(\alpha)$.

\begin{figure}[th]
    \includegraphics{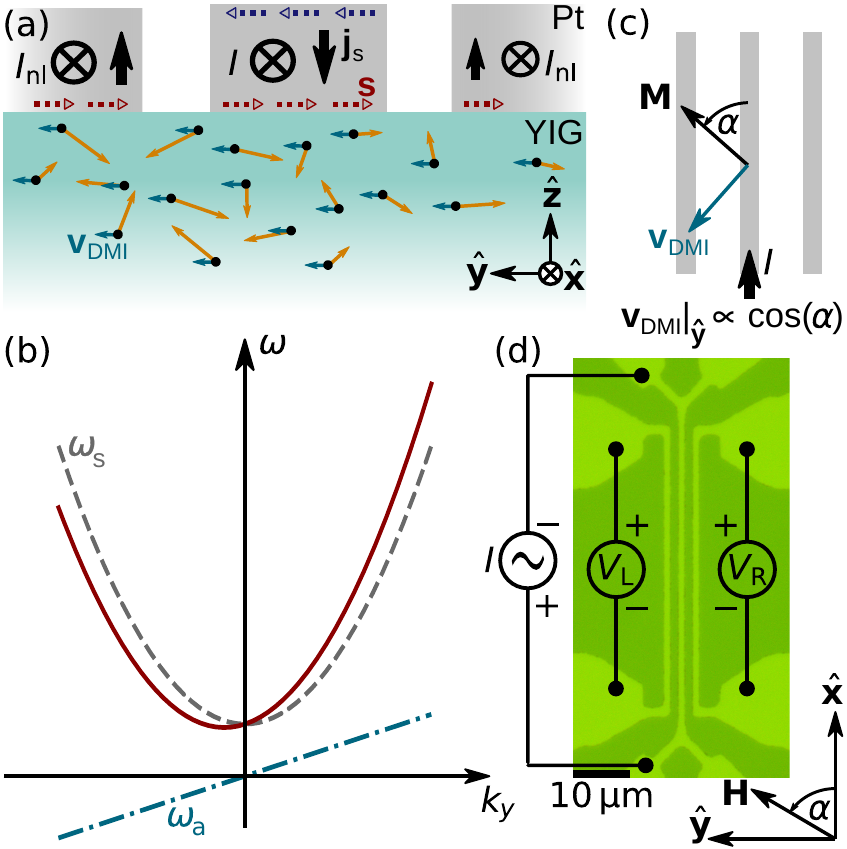}
    \caption{\label{fig:f1}
    (a) Driving an electrical current $I$ in the central wire generates a spin accumulation (open dashed arrows) via the spin Hall effect so that magnons can be excited in the YIG layer. 
    In the presence of DMI, a finite drift velocity $\mathbf{v}_\mr{DMI}$ (blue arrows) is superposed onto the random diffusive magnon motion (orange arrows). 
    Thus, the higher number of magnons propagating to the left Pt wire compared to the right one leads to a difference in magnon population and in turn a different inverse spin Hall current $I_\mr{nl}$.
    (b) The magnon dispersion relation in a system with finite DMI (red line) has a symmetric parabolic dispersion (gray dashed line), which leads to random diffusion and an asymmetric (linear) contribution due to DMI (blue dash-dotted line) which results in unidirectional drift.
    (c) For interfacial DMI, the orientation of the drift velocity is orthogonal to the in-plane magnetization, providing control over the drift velocity along the magnon propagation direction $\hat{\mathbf{y}}$ [see \eref{vdrift}].
    (d) Optical micrograph of a representative three wire device, and sketches of the electrical wiring and coordinate system.
    }
\end{figure}

\emph{Drift-diffusion model.} We first generalize the theory of spin transport driven by the magnon chemical potential in a ferromagnet~\cite{Cornelissen2016a} to include inversion asymmetry in the system. 
As introduced above, the magnon dispersion bears a contribution odd in wavevector~\cite{Udvardi2009, Moon2013, Nembach2015} leading to a drift-like term in the transport equation~\cite{Zutic2004}. 
Separating the symmetric and asymmetric contributions $\omega(\mathbf{k}) = \omega_{\mathrm{s}}(\mathbf{k}) + \omega_{\mathrm{a}}(\mathbf{k})$ and considering $\omega_{\mathrm{a}}(\mathbf{k}) = \mathbf{v}_{\mathrm{DMI}}\cdot \mathbf{k}$ (see \fref{f1}b) we limit our considerations to the first order in the drift velocity $\mathbf{v}_{\mathrm{DMI}}$ and thereby assume its effect to be small.
As we focus on spin transport driven by the magnon chemical potential~\cite{Cornelissen2015,Cornelissen2016a}, we assume a constant temperature $T$ throughout the system.
The spatially resolved magnon density can be expressed as~\cite{Kittel2004}:
\begin{align}
    n(\mathbf{r})  & =  \int \frac{d^3 k}{(2 \pi)^3} n_\mr{B}(\hbar \omega (\mathbf{k}) - \mu (\mathbf{r})), \nonumber \\
    & \approx \int \frac{d^3 k}{(2 \pi)^3} \left[  n_\mr{B}(\hbar \omega_{\mathrm{s}})  +  \left ( \left. - \frac{\partial n_\mr{B}(\epsilon)}{\partial \epsilon} \right|_{\epsilon = \hbar \omega_{\mathrm{s}}} \right) \mu(\mathbf{r}) \right], \nonumber \\
 & \equiv n_0 + \chi \mu (\mathbf{r}), \label{eq:density}
\end{align}
where $n_\mr{B}(\epsilon) \equiv 1/(\exp(\epsilon/k_\mr{B} T) - 1)$ is the Bose distribution function determining the magnon population, $k_B$ the Boltzmann constant, $\mu(\mathbf{r})$ the magnon chemical potential, and $n_0$ the equilibrium magnon density.
As per Eq.~\eqref{eq:density}, the contribution of $\omega_\mr{a}$ to the dispersion does not affect the magnon density.
The former vanishes when integrating over all $\mathbf{k}$ as $\omega_{\mathrm{a}}$ is odd in $\mathbf{k}$.
The magnon current density, however, now includes a drift contribution stemming from $\omega_{\mathrm{a}}$ in addition to the diffusive component considered in previous studies~\cite{Cornelissen2015,Cornelissen2016a}:
\begin{align}
    \mathbf{j}_{m} & =  - D_m \mathbf{\nabla} n(\mathbf{r}) +  \mathbf{v}_{\mathrm{DMI}} n(\mathbf{r}), \label{eq:current}
\end{align}
where $D_m$ is the magnon diffusion constant~\cite{Cornelissen2016a}, \rs{which is independent of the DMI up to first order in $\omega_\mr{a}$.}
Employing Eqs.~\eqref{eq:density} and \eqref{eq:current} in the continuity equation for magnons and parametrizing the decay of nonequilibrium magnons via a relaxation time $\tau_m$, we obtain the transport equation:
\begin{align}\label{eq:transport}
\frac{\partial \mu}{\partial t} - D_m \nabla^2 \mu + \mathbf{v}_{\mathrm{DMI}} \cdot \mathbf{\nabla} \mu & = - \frac{\mu}{\tau_m}.
\end{align}
The term $\mathbf{v}_\mr{DMI} \cdot \mathbf{\nabla} \mu$ represents drift and is the only addition as compared to the analogous description for inversion-symmetric systems~\cite{Cornelissen2016a}. 

For our quasi-dc nonlocal transport experiments (see \fref{f1}), we solve Eq.~\eqref{eq:transport} above in steady state and in the one-dimensional limit, i.e.,~assuming the chemical potential to depend only on the $y$ coordinate. Substituting the ansatz $\mu \sim \exp (- y/l_p)$ for the magnon chemical potential, we obtain:
\begin{align}
    D_m \tau_m + v_{\mathrm{DMI}} \cos(\alpha)  \tau_m l_p & = l_p^2. 
\end{align} 
The magnon propagation length $l_p$ is thus obtained as:
\begin{align}\label{eq:lp}
    l_{p} & \approx \lambda_0 + \frac{v_{\mathrm{DMI}} \tau_m \cos(\alpha)}{2},
\end{align}
where $\lambda_0 \equiv \sqrt{D_m \tau_m}$ is the magnon diffusion length. 
Thus, we see that the magnon propagation length bears a contribution from drift which can be positive or negative and is controllable via the orientation of the magnetic field given by $\alpha$ for the case at hand. Within a simplified model, the nonlocal resistance (\ie the nonlocal voltage due to magnon propagation divided by the injector current) detected along a given Pt detector wire is proportional to the magnon chemical potential~\cite{Cornelissen2016a}:
\begin{align}\label{eq:detv}
    R_{\mathrm{nl}} & \propto  e^{- \frac{d_\mr{nl}}{l_p}} \sin^2(\alpha) , \nonumber \\
    & \approx e^{- \frac{d_\mr{nl}}{\lambda_0}} \sin^2(\alpha) \left( 1 + \frac{d_\mr{nl} v_{\mathrm{DMI}} \tau_m \cos(\alpha) }{2 \lambda_0^2} \right) ,
\end{align}
where $d_\mr{nl}$ is the injector-detector distance. 
\rs{This expression is valid in the limit of $v_\mr{DMI}\tau_m/2\lambda_0 \ll 1$, \ie when the DMI contribution to the nonlocal transport is small compared to its diffusive counterpart}. 
The $\sin^2(\alpha)$ dependence arises due to the spin injection and detection via the (inverse) spin Hall effect~\cite{Cornelissen2015, Zhang2012, Zhang2012a}.

We conclude that \rs{the $\sin^2(\alpha)$ angular dependence of the nonlocal signal will be altered by the finite drift velocity.}
Additionally, we note that due to the inversion asymmetric nature of the DMI the second term in \eref{detv} will change sign when magnon transport along the $-\hat{\mathbf{y}}$ direction is considered.
We thus expect that \eref{detv} describes the resistance of the left detector and the drift response on the right detector will be opposite.


\emph{Experimental Details.} YIG films with \rs{thickness $t_\mr{YIG} = 30 - \SI{100}{\nano\meter}$} were grown onto (111)-oriented \ch{Gd3Ga5O12} (GGG) substrates from a stoichiometric \ch{Y3Fe5O12} target with a radio-frequency sputtering system with a base pressure better than \SI{1e-5}{\pascal}. 
The growth temperature and the Ar pressure during the growth was \SI{800}{\celsius} and \SI{1.3}{\pascal}, respectively.
After deposition, the samples were annealed for \SI{30}{\minute} at the deposition temperature and finally cooled back to room temperature in vacuum.
A \SI{4}{\nano\meter} thick Pt layer was subsequently deposited in the same chamber at room temperature using dc~sputtering in Ar atmosphere at a pressure of \SI{0.4}{\pascal}. 
For reference, a YIG/Pt heterostructure was prepared on a \SI{150}{\nano\meter} thick YIG film grown via liquid phase epitaxy on GGG (see Ref.~\cite{Schlitz2019a} for details).
\rs{The formation of a spurious \ch{Gd3Fe5O12} layer at the GGG/YIG interface~\cite{GomezPerez2018} was excluded via local energy dispersive X-ray spectroscopy (see Ref.~\cite{SMDMI}).}
The Pt films were then patterned into several devices with three parallel wires, having center-to-center separations of \SI{1.5}{\micro\meter} to \SI{8}{\micro\meter} either by optical or electron beam lithography and subsequent Ar ion milling.
All investigated devices have a wire width of \SI{1}{\micro\meter} and a detector length of \SI{50}{\micro\meter} or \SI{120}{\micro\meter}. 

The measurements were performed in a room temperature electromagnet setup.
A sinusoidal current with peak amplitude $I_0 = \SI{1}{\milli\ampere}$ and frequency $f = \SI{10}{\hertz}$ was applied to the central wire while \SI{2}{\second} to \SI{10}{\second}-long time traces of the voltages on the left and right Pt wire were simultaneously acquired and subsequently demodulated to obtain the first harmonic signal $R_{1\omega} = V_{1\omega}/I_0$ (see also Ref.~\cite{Garello2013}).
\rs{This method allows for disentangling the electronic (first harmonic) and thermal (second harmonic) contributions to the measured voltage~\cite{Cornelissen2015, SMDMI}.}
An optical micrograph including the sketched electrical contacts and the definition of the rotation angle is presented in \fref{f1}d.

\begin{figure}[th]
    \includegraphics{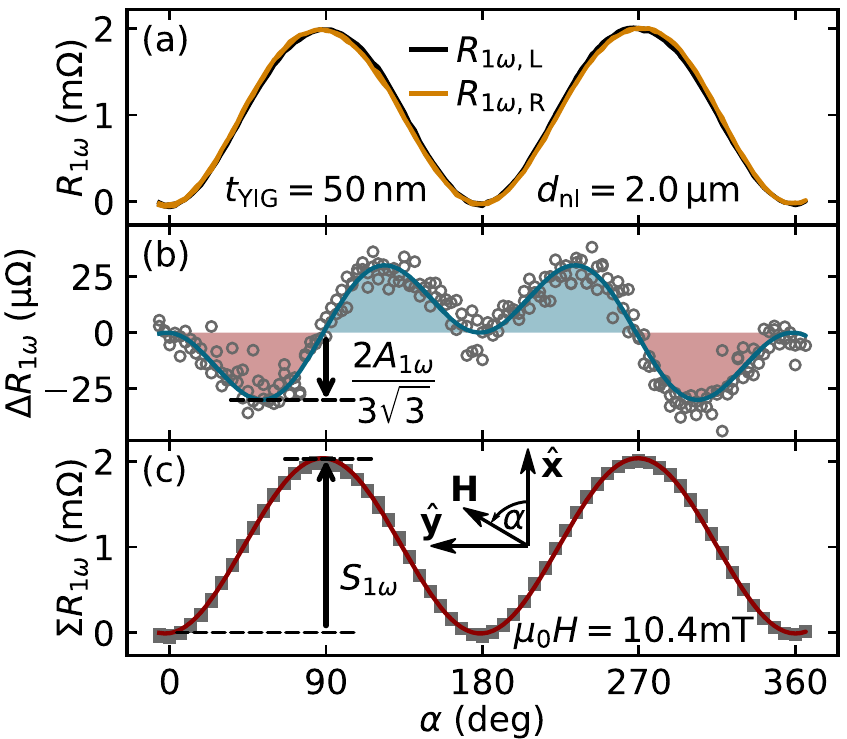}
    \caption{\label{fig:f2}
        (a) The nonlocal resistance $R_{1\omega}$ of the left (black line) and right (orange line) wire shows a different dependence on magnetic field orientation $\alpha$. 
	$R_{1\omega, \mr{R}}$ has been rescaled by $\SI{0.4}{\percent}$ to scale with $R_{1\omega, \mr{L}}$, a difference arising from device imperfections.
	Additionally, a constant offset arising from inductive or capacitive coupling has been removed from both curves~\cite{Cornelissen2015}.
        (b) The difference between the two curves in panel (a) (grey circles) corresponds to a difference in magnons transported to the left and right Pt wire.
        The blue curve represents a $A_{1\omega} \sin^2(\alpha)\cos(\alpha)$ angular dependence. 
	The shaded blue (red) regions correspond to enhanced (reduced) magnon transport towards the left wire.
        (c) The average of the two curves in panel (a) (solid gray squares). 
	The thickness of the YIG film is $t_\mr{YIG} = \SI{50}{\nano\meter}$.
        The red curve represents a $S_{1\omega}\sin^2(\alpha)$ dependence, which is expected in the absence of a drift contribution. 
    }
\end{figure}


\emph{Experimental Results.} \Fref{f2}a shows a representative set of nonlocal magnon transport data taken in a device with \SI{50}{\nano\meter} thick YIG and a wire separation of $d_\mr{nl} = \SI{2.0}{\micro\meter}$.
For purely diffusive transport, $R_{1\omega}$ should exhibit a $\sin^2(\alpha)$ dependence on the in-plane angle of the magnetic field $\alpha$ [\cf \eref{detv}]~\cite{Cornelissen2015, Zhang2012, Zhang2012a}.
In contrast, we observe that $R_{1\omega}$ shows a small distortion from the expected shape, which depends on the angle of the magnetic field and differs between the left and right wire.
To isolate this asymmetric contribution, we plot the difference between the nonlocal resistance of the two wire $\Delta R_{1\omega} = (R_{1\omega\mr{, L}} - R_{1\omega\mr{, R}})/2$ in \fref{f2}b.
$\Delta R_{1\omega}$ reflects the difference in the amount of magnons transported to the left and right wire and corresponds to the second term in \eref{detv}.
We indeed find that its angular dependence can be well described by $\Delta R_{1\omega} = A_{1\omega} \cos(\alpha)\sin^2(\alpha)$, where $A_{1\omega}$ quantifies the amplitude of the directional drift contribution.
Note that for purely diffusive magnon transport, \ie when $v_\mr{DMI} = 0$, $A_{1\omega}$ vanishes~\cite{Cornelissen2015}.
We thus conclude that a finite magnon drift contribution to the nonlocal magnon transport is present in our heterostructures. 
In particular, magnons are transported more (less) efficiently to the left Pt wire when $\alpha$ is between \SI{90}{\degree} (\SI{270}{\degree}) and \SI{270}{\degree} (\SI{90}{\degree}), which is represented by the blue (red) shaded areas in \fref{f2}b.
The average signal $\Sigma R_{1\omega} = (R_{1\omega\mr{, L}} + R_{1\omega\mr{, R}})/2$, which includes only the diffusive contribution, is shown in \fref{f2}c and closely resembles a $\sin^2(\alpha)$ with amplitude $S_{1\omega}$ as expected~\cite{Cornelissen2015} [first term in \eref{detv}].
\rs{The current dependence of $S_{1\omega}$ and $A_{1\omega}$ as well as a possible effect of the DMI on thermally excited magnon transport is discussed in Ref.~\cite{SMDMI}.}

\begin{figure}[th]
    \includegraphics{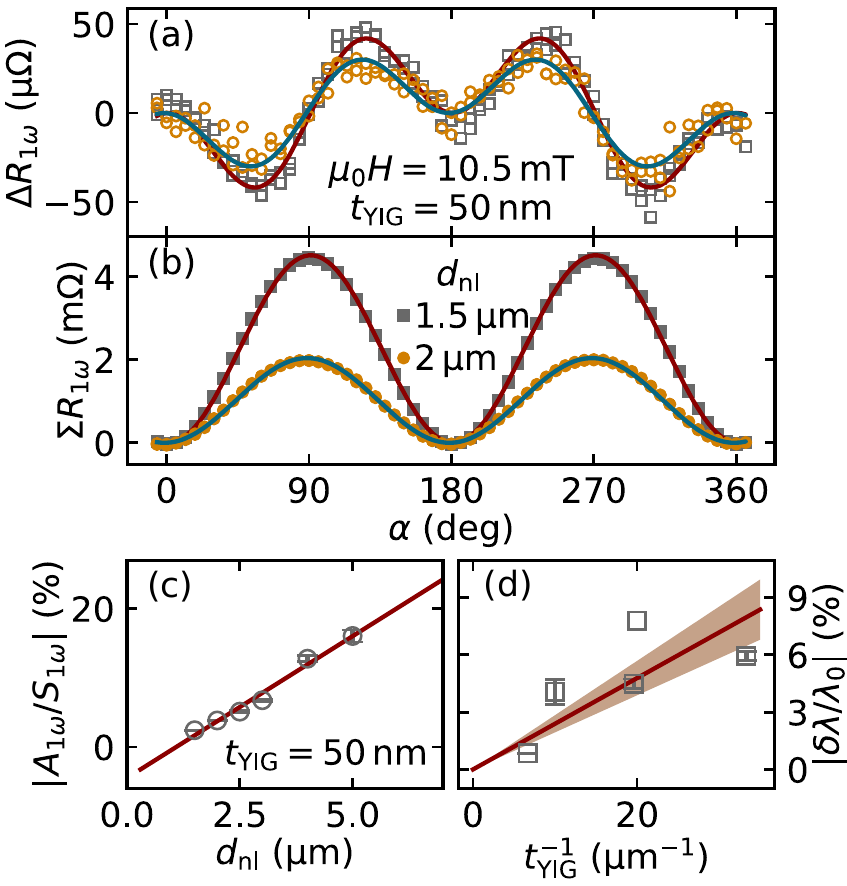}
    \caption{\label{fig:f3}
    \rs{(a) Dependence of $\Delta R_{1\omega}$ and (b) $\Sigma R_{1\omega}$ on the wire separation $d_\mr{nl}$.
	The gray (orange) symbols correspond to $d_\mr{nl} = \SI{1.5}{\micro\meter}$  ($\SI{2.0}{\micro\meter}$) and} the solid lines are fits of the respective angular dependencies.
        (c) $A_{1\omega}/S_{1\omega}$ increases linearly with the distance $d_\mr{nl}$ between the central wire and the wires to the left and right (red line).
        The error bars are estimated from fits to the angular dependency of $\Delta R_{1\omega}$ and $\Sigma R_{1\omega}$.
        (d) The normalized directional propagation length as a function of the inverse of \rs{$t_\mr{YIG}$}.
        The error for $\delta\lambda/\lambda_0$ is evaluated from the fits to the distance dependence of $A_{1\omega}/S_{1\omega}$ and $S_{1\omega}$ and the shaded red area corresponds to the error of the linear fit.
    }
\end{figure}


We will now verify that the experimentally observed magnon drift contribution follows the theoretical prediction [\eref{detv}].
To that end, we characterize the spatial decay of the drift contribution $\Delta R_{1\omega}$ and the diffusive contribution $\Sigma R_{1\omega}$ corresponding to the first and second term in \eref{detv}, respectively.
\Fref{f3}a and b show representative data of $\Delta R_{1\omega}$ and $\Sigma R_{1\omega}$ for devices with $d_\mr{nl} = \SI{1.5}{\micro\meter}$ and $d_\mr{nl} = \SI{2.0}{\micro\meter}$.
While the amplitude of $\Delta R_{1\omega}$ decays to $\SI{70}{\percent}$, $\Sigma R_{1\omega}$ drops to \SI{44}{\percent} upon increasing $d_\mr{nl}$.
We conclude that the antisymmetric (drift) contribution to the magnon transport (with amplitude $A_{1\omega}$) decays significantly slower than its diffusive counterpart (with amplitude $S_{1\omega}$) as predicted by our theory [\cf \eref{detv}].
\Fref{f3}c displays the ratio of the antisymmetric and symmetric contributions $A_{1\omega}/S_{1\omega}$, which evolves linearly with distance as expected from \eref{detv} and reaches a maximum value of up to \SI{16}{\percent} for the largest distance $d_\mr{nl} = \SI{5}{\micro\meter}$.
\rs{The} slope of the linear fit of $A_{1\omega}/S_{1\omega}$ is directly proportional to $\delta \lambda/\lambda_0^2$, where $\delta \lambda \equiv v_\mr{DMI}\tau_m/2$ [see \eref{detv}].
$\lambda_0$ is determined by fitting the exponential decay of $S_{1\omega}$ with $d_\mr{nl}$ for each film~\cite{Cornelissen2015}.
\rs{We find $\SI{400}{\nano\meter} \lesssim \lambda_0 \lesssim \SI{2}{\micro\meter}$ in good agreement with the range reported in the literature for nonlocal magnon transport in YIG/Pt heterostructures~\cite{Goennenwein2015, Cornelissen2015, Wimmer2019}}
The presence of a directional contribution to the nonlocal transport and its dependence on distance shown in \fref{f3}c verify the theoretical predictions and thus further corroborate the origin of the former to be magnon drift currents.

Finally, to pinpoint the origin of the magnon drift velocity, we investigate the dependence of $\delta \lambda/\lambda_0$ on \rs{$t_\mr{YIG}$}.
\Fref{f3}d shows that for decreasing \rs{$t_\mr{YIG}$}, the relative drift contribution increases, suggesting that the magnon drift velocity originates from interfacial DMI.
This conclusion is consistent with recent reports of interfacial DMI at the GGG/YIG interface, inferred from Brillouin light scattering (BLS) and spin wave spectroscopy measurements~\cite{Wang2020}, as well as from current-induced domain wall motion experiments in (S)GGG/TmIG/Pt heterostructures~\cite{Velez2019a, Avci2019, Ding2019}.
To provide a rough estimate of the DMI strength, we take the logarithmic mean $\tau_m \sim \SI{10}{\nano\second}$ from the reported range $\SI{1}{\nano\second} \lesssim \tau_m \lesssim \SI{100}{\nano\second}$~\cite{Cornelissen2016a, Flebus2017}.
From $\delta\lambda/\lambda_0 \sim -\SI{6}{\percent}$ we can thus estimate $v_\mr{DMI} \sim -\SI{5}{\meter/\second}$ for the sample with \SI{30}{\nano\meter} thick YIG, and consequently find $D \sim -\SI{2}{\micro\joule\per\meter^2}$ [see \eref{vdrift}].
\rs{Considering that the interfacial DMI scales inversely with thickness, we find that $D$ agrees within a factor two with recent measurements of GGG/YIG and SGGG/TmIG heterostructures~\cite{Wang2020, Velez2019a}.}
Since the main transport channel is not covered by the Pt layer, we conclude that the DMI originates from the GGG/YIG interface, in agreement with other studies~\cite{Velez2019a,Ding2019,Avci2019,Wang2020}.


\emph{Summary.} We have demonstrated a magnon drift contribution to the diffusive nonlocal magnon transport in all-electrical magnon transport devices.
The magnon drift velocity can be controlled by the orientation of the magnetic field and gives rise to a characteristic angular dependence.
We extended the theory of magnon transport driven by a magnon chemical potential by including magnon drift and found excellent agreement with experiment. 
Finally, we studied the thickness dependence of the drift contribution, revealing the interfacial origin of the DMI in the YIG/GGG heterostructures.
The DMI effectively gives rise to a directional driving force on the magnons, which provides an additional handle to tune the magnon transport properties in nonlocal devices.
Magnon drift currents can be realized via various other inversion symmetry-breaking mechanisms and are intrinsic to the materials, paving the way to their application in future devices.

\nocite{Meyer2017a}

\begin{acknowledgments}
    We acknowledge financial support by the Swiss National Science Foundation (SNSF) via projects no.~198642 and 20020\_172775, 
    by the Deutsche Forschungsgemeinschaft via SFB 1143 (project no.~C08), 
    by the ETH Z{\"u}rich through the Career Seed Grant SEED-20 19-2, 
    by the Research Council of Norway through its Centers of Excellence funding scheme, project 262633, ``QuSpin'' 
	and through the W{\"u}rzburg-Dresden Cluster of Excellence on Complexity and Topology in Quantum Matter - ct.qmat (EXC 2147, project no.~39085490). 
        \rs{We also acknowledge the Dresden Center for Nanoanalysis (DCN) at the Technische Universit{\"a}t Dresden and the support of Alexander Tahn and Darius Pohl.}
\end{acknowledgments}

\bibliography{190618_bibliography.bib}

\appendix{}

\end{document}